\documentstyle[prb,aps,floats,epsfig,eqsecnum,amssymb]{revtex}
\begin{document}
\twocolumn[\hsize\textwidth\columnwidth\hsize\csname@twocolumnfalse
\endcsname
\title{Static and dynamic structure factors in the Haldane phase \\
  of the bilinear$-$biquadratic spin$-$1 chain}
\author{Andreas Schmitt and Karl-Heinz M\"utter}
\address{Department of Physics, University of Wuppertal, 
                D-42097 Wuppertal, Germany}
\author{Michael Karbach\cite{aaa}, Yongmin Yu, and Gerhard M\"uller}
\address{Department of Physics, The University of Rhode Island, 
                Kingston, Rhode Island 02881-0817}
\date{\today}
\maketitle
%
%
%
%
\begin{abstract}
  The excitation spectra of the $T=0$ dynamic structure factors for the spin,
  dimer, and trimer fluctuation operators as well as for the newly defined
  center fluctuation operator in the one-dimensional $S=1$ Heisenberg model with
  isotropic bilinear $(J\cos\theta)$ and biquadratic $(J\sin\theta)$ exchange
  are investigated via the recursion method for systems with up to $N=18$ sites
  over the predicted range, $-\pi/4<\theta\lesssim\pi/4$, of the topologically
  ordered Haldane phase. The four static and dynamic structure factors probe the
  ordering tendencies in the various coupling regimes and the elementary and
  composite excitations which dominate the $T=0$ dynamics. At $\theta =
  \arctan\frac{1}{3}$ (VBS point), the dynamically relevant spectra in the
  invariant subspaces with total spin $S_T = 0,1,2$ are dominated by a branch of
  magnon states $(S_T = 1)$, by continua of two-magnon scattering states $(S_T =
  0,1,2)$, and by discrete branches of two-magnon bound states with positive
  interaction energy $(S_T = 0,2)$. The dimer and trimer spectra at $q=\pi$ are
  found to consist of single modes with $N$-independent excitation energies
  $\omega_\lambda^D/|e_0|=5$ and $\omega_\lambda^T/|e_0|=6$, where $e_0=E_0/N$
  is the ground-state energy per site. The basic structure of the dynamically
  relevant excitation spectrum remains the same over a substantial parameter
  range within the Haldane phase. At the transition to the dimerized phase
  ($\theta=-\pi/4$), the two-magnon excitations turn into two-spinon
  excitations.
\end{abstract}
\twocolumn
]
%
%
\section{Introduction}\label{sec:intro}
%
%
The zero-temperature phase diagram of the one-dimensional (1D) $S=1$ Heisenberg
model with bilinear and biquadratic exchange,
\begin{equation}\label{H}
       H_\theta = J\sum_{l=1}^N \left[ 
         \cos \theta ({\bf S}_l \cdot {\bf S}_{l+1}) + 
         \sin \theta \left( {\bf S}_l \cdot {\bf S}_{l+1} \right)^2\right],
\end{equation}
$-\pi < \theta \leq \pi$, as is widely accepted
today,\cite{FS91,FS95,BXG95,SJG96} consists of a phase with ferromagnetic
long-range order ($\theta<-3\pi/4$ or $\theta>\pi/2$), a phase with dimer
long-range order ($-3\pi/4<\theta<-\pi/4$), the Haldane phase with hidden
topological long-range order ($-\pi/4<\theta<\pi/4$), and a somewhat obscure
``trimerized'' phase ($\pi/4<\theta<\pi/2$). $H_\theta$ is Bethe-ansatz solvable
at $\theta=-\pi/4$,\cite{Takh82} and at
$\theta=\pi/4$,\cite{Uimi70} and for part of the eigenvalue
spectrum also at $\theta=-\pi/2$.\cite{Park87} Within the Haldane phase, at the
parameter value $\theta_{VBS}=\arctan\frac{1}{3}\simeq 0.1024\pi$, the
ground-state wave function of $H_\theta$ is a realization of the (explicitly
known) valence-bond solid (VBS) wave function.\cite{AKLT87,AKLT88,AAH88}

In spite of numerous theoretical and computational studies of this model system,
there are still many blank spots on the map, especially with respect to
dynamical properties. A panoramic view of the various ordering tendencies and of
the dominant quantum fluctuations at $T=0$ can be gained from a study of the
dynamic structure factors,
\begin{equation}\label{SAAqw}
  S_{AA}(q,\omega) \equiv 
  \int\limits_{-\infty}^{+\infty} dt\,e^{i\omega t} 
  \langle F_q^A(t)F_q^{A^\dagger} \rangle,
\end{equation} 
for a set of fluctuation operators $F_q^A$. Among them should be the operators
which, for specific wave numbers $q$, describe the known or suspected order
parameters. Further fluctuation operators may be chosen according to specific
symmetry properties of $H_\theta$, which entail special selection rules.

Each dynamic structure factor calculated for the same parameter value $\theta$
has its own dynamically relevant excitation spectrum and its own spectral-weight
distribution. Each fluctuation operator acts like a filter on the complete
eigenvalue spectrum, and the associated dynamic structure factor highlights a
particular subset of excitations. Looking at the $T=0$ dynamics of $H_\theta$
through a series of such filters reveals many interesting features that are not
readily accessible by any other means. The composite structure of parts of the
excitation spectrum and the nature of the underlying elementary excitations, for
example, may be recognized only if observed through the right set of filters.

For some questions, it is useful to investigate the corresponding static
structure factors,
\begin{equation}\label{SAAq}
        S_{AA}(q) \equiv \langle F_q^A F_q^{A^\dagger} \rangle = 
        \int\limits_{-\infty}^{+\infty} \frac{d\omega}{2\pi} S_{AA}(q,\omega),
\end{equation}
which are more likely amenable to a finite-size scaling analysis. All these
quantities can be computed from the finite-size ground-state wave function,
$S_{AA}(q)$ directly as an expectation value and $S_{AA}(q,\omega)$ indirectly
via the recursion method.\cite{VM94,FKMW95}

The fluctuation operators used in this study and their relation to different
order parameters are introduced in Sec.~\ref{sec:flucop}. The static and dynamic
structure factors which probe the different kinds of fluctuations are
investigated in Sec.~\ref{sec:vbs} for the VBS point and in Sec.~\ref{sec:hal}
for the surrounding parts of the Haldane phase. The continuation of this study
for the parameter values at the two critical points $\theta=\pm\pi/4$ and in the
phases beyond these points will be reported elsewhere.\cite{note8}
%
%
\section{Fluctuation operators and \\ order parameters}
\label{sec:flucop}
%
%
For the study of the model system (\ref{H}) we introduce four different
fluctuation operators of the general form
\begin{equation}\label{FqA}
        F_q^A \equiv \frac{1}{\sqrt{N}}\sum_{l=1}^Ne^{iql}A_l,
\end{equation}
where the operator $A_l$ acts locally at lattice site $l$ and (in some cases)
also on one or two neighboring sites.

(i) The {\it spin} fluctuations are probed by the operator $F_q^S$ with
\begin{equation}\label{Sl}
        S_l\equiv S_l^3 = \left(
                \begin{array}{ccc}
                        1 & 0 & 0 \\
                        0 & 0 & 0 \\
                        0 & 0 &-1
                \end{array}     
                        \right)_l .
\end{equation}
For $q=0$ they represent ferromagnetic order-parameter fluctuations and for
$q=\pi$ N\'eel order-parameter fluctuations. Ferromagnetic long-range order
does exist in this model, but N\'eel long-range order does not. The N\'eel
order-parameter fluctuations are expected to be strongest at the critical point
$\theta=-\pi/4$, which marks the disappearance of topological long-range order,
the only point in the phase diagram where the $q=\pi$ spin excitations are known
to be gapless.

(ii) The {\it dimer} fluctuations are characterized by the operator $F_q^D$
with\cite{note5}  
\begin{equation}\label{Dl}
        D_l \equiv {\bf S}_l \cdot {\bf S}_{l+1} -
        \langle {\bf S}_l \cdot {\bf S}_{l+1} \rangle.
\end{equation}
The dimer order-parameter fluctuations, probed by $F_\pi^D$, are expected to be
strongest at the same critical point, $\theta=-\pi/4$, which also marks the
onset of dimer long-range order.

(iii) For the study of {\it trimer} fluctuations, we use the operator $F_q^T$
with
\begin{equation}\label{Tl}
        T_l \equiv P_l^T - \langle P_l^T \rangle
\end{equation}
constructed from projection operators 
\begin{equation}
        P_l^T \equiv |[l,l+1,l+2]\rangle \langle [l,l+1,l+2] |
\end{equation}
onto local trimer states
\begin{eqnarray}\label{state123}
        |[1,2,3]\rangle \equiv \frac{1}{\sqrt{6}} 
&&  
        \left(|+0-\rangle + |0-+\rangle + |-+0\rangle \right.
\nonumber \\ ~~  - &&
        \left.|-0+\rangle - |0+-\rangle - |+-0\rangle \right),
\label{lts}
\end{eqnarray}
which are completely antisymmetric states with total spin
$S_T\!=\!0$.\cite{NT91} The state (\ref{state123}) is, in fact, the
(non-degenerate) ground state of $H_\theta$ with $N=3$ for
$\arctan\frac{1}{3}\leq\theta\leq\pi/2$. This observation was interpreted as
suggesting that a phase with trimer long-range order might exist for
$N\to\infty$. The trimer order-parameter fluctuations are probed by
$F_{2\pi/3}^T$.

(iv) Our finite-$N$ data indicate that in the same parameter range,
$\pi/4\leq\theta\leq\pi/2$, where the trimer order-parameter fluctuations are
strong, the spin fluctuations are significantly stronger, and the fluctuations
of a modified spin operator, which tunes into existing period-three patterns of
local up-zero-down $(+,0,-)$ or down-zero-up $(-,0,+)$ spin states are even
stronger.  The {\it center } fluctuation operator $F_q^Z$ is constructed from
the matrices
\begin{eqnarray} Z_l\equiv&& Z_l^3 =
        \left( \begin{array}{ccc} 
                        e^{i2\pi/3} & 0 & 0 \\ 
                        0 & 1 & 0 \\         
                        0 & 0 & e^{-i2\pi/3} 
                        \end{array} \right)_l \nonumber \\
        =&& {\bf 1} + i \frac{\sqrt{3}}{2}S_l^z -\frac{3}{2}(S_l^z)^2, 
\label{Zl}
\end{eqnarray} 
which is an element of the SU(3) center.\cite{note1} The center
order-parameter has the same wave number, $q=2\pi/3$, as the trimer
order-parameter. At $\theta=\pi/4$ the excitation spectrum of $H_\theta$ is
rigorously known to be gapless for this wave number.

In Appendix A we discuss the static spin, dimer, trimer, and center
correlation functions in special states constructed such as to reflect pure
N\'eel, dimer, trimer, or center long-range order.

All four fluctuation operators $F_q^A$ are invariant under (discrete)
translations in real space and under (continuous) rotations about the 3-axis in
spin space. The resulting selection rules for the excited states that may
contribute to any of the four dynamic structure factors $S_{AA}(q,\omega)$ are
$\Delta k=q$ for the wave number and $\Delta S_T^3=0$ for the 3-component of the
total spin.

Only two of the fluctuation operators are fully rotationally invariant,
$[F_q^D,S_T^i]=[F_q^T,S_T^i]=0$ for $i=1,2,3$ and arbitrary $q$. This produces
the additional selection rules $\Delta S_T=0$ for the magnitude of the total
spin in the dynamic dimer and trimer structure factors.  The corresponding
selection rules for the dynamic spin and center structure factors are $\Delta
S_T=0,1$ and $\Delta S_T=0,1,2$, respectively, with the further restriction that
transitions between singlets ($S_T=0$ states) are prohibited.

In the non-ferromagnetic parameter range ($-3\pi/4<\theta<\pi/2$), where the
finite-$N$ ground state is a non-degenerate singlet, the dynamic structure
factors $S_{DD}(q,\omega)$ and $S_{TT}(q,\omega)$ thus couple exclusively to the
$S_T=0$ excitation spectrum, and $S_{SS}(q,\omega)$ exclusively to the $S_T=1$
excitation spectrum, whereas $S_{ZZ}(q,\omega)$ couples to the $S_T=1$ and
$S_T=2$ spectra.

To calculate $S_{AA}(q,\omega)$ we employ the recursion
method\cite{Hayd80,VM94} in combination with a finite-size
continued-fraction analysis.\cite{FKMW95} The recursion algorithm in the present
context is based on an orthogonal expansion of the wave function
$|\Psi_q^A(t)\rangle = F_q^A(-t)|G\rangle$, where $|G\rangle$ is the finite-$N$
ground state. It produces a double sequence of continued-fraction coefficients
$\{a_k(q),b_k^2(q)\}$ for the function
\begin{equation}\label{d0AA}
  d_0^{AA}(q,\zeta) = 
  {\displaystyle {\frac{i}{\zeta-a_0(q)-
        {\displaystyle{\frac{b_1^2(q)}{\zeta-a_1(q)-
              {\displaystyle{\frac{b_2^2(q)}{\zeta-a_2(q)-\ldots}}}}}}}}},
\end{equation}  
which is the Laplace transform of the correlation function
$\langle F_q^A(t)F_q^{A^\dagger}\rangle$, and from which the dynamic structure
factor (\ref{SAAqw}) can be directly recovered via the relation 
\begin{equation}
        S_{AA}(q,\omega) = 2\langle F_q^A F_q^{A^\dagger}\rangle 
                        \lim_{\epsilon\to 0} \Re [d_0^{AA}(q,w+i\epsilon)].
\end{equation}
The finite-size continued-fraction analysis expresses the dynamic structure
factor in the form
\begin{equation}
      S_{AA}(q,\omega)= \sum_\lambda W_\lambda^A \delta(\omega-\omega_\lambda^A),
\end{equation}
where the sum runs over all the dynamically relevant excitations
$|\lambda\rangle$ with frequency $\omega^A_\lambda$ and spectral weight
$W_\lambda^A=2\pi|\langle G|F_q^A|\lambda\rangle|^2$. For the excitations which
carry the bulk of the spectral weight both $\omega_\lambda^A$ and $W_\lambda^A$
can be extracted quite accurately from a finite number of
continued-fraction coefficients.

We begin our study of the static and dynamic structure factors of $H_\theta$ at
the VBS point, where each of the four ordering tendencies introduced previously
(N\'eel, dimer, trimer, center) is imperceptibly weak, and where some relevant
static quantities are exactly known.
%
%
\section{VBS State}\label{sec:vbs}
%
%
%
%
\subsection{Static structure factors}\label{sec:ssf}
%
%
The ground-state wave function of $H_\theta$ at $\theta=\arctan\frac{1}{3}$ can
be constructed for arbitrary $N$ as follows:\cite{AKLT87,AKLT88,AAH88} The spin
1 at each lattice site is expressed as a spin-1/2 pair in a triplet state. The
singlet-pair forming valence bond involves one fictitious spin 1/2 from each of
two neighboring lattice sites. The VBS state can then be regarded as a chain of
valence bonds linking successive spin-1/2 pairs in this manner.

The spin,\cite{AKLT87,AAH88} dimer,\cite{Deis92} trimer, and center
order-parameter correlation functions and the associated static structure
factors can be determined exactly:

\begin{mathletters}
\begin{eqnarray}
        \langle S_l^z S_{l+n}^z\rangle 
&=& 
        \frac{2}{3}\left(\frac{2(-1)^n}{3^{|n|}}-\delta_{n0}\right),
\\
        S_{SS}(q)
&=&
        \frac{2(1-\cos q)}{5+3 \cos q};
\end{eqnarray}
\end{mathletters}
\begin{equation}\nonumber
        \langle D_l D_{l+n}\rangle = \frac{2}{9} \delta_{n0},
        \quad
        S_{DD}(q) = \frac{2}{9};
\end{equation}
\begin{mathletters}
\begin{eqnarray}
        \langle T_l T_{l+n} \rangle &=& \frac{\sqrt{12}}{20}\delta_{0l} + 
    \frac{1}{40}(\delta_{1l}+\delta_{-1l}), \\
    S_{TT}(q) &=& \frac{1}{20}\left(\sqrt{12}+\cos q \right);
\end{eqnarray}
\end{mathletters}
\begin{equation}
        \langle Z_l^\dagger Z_{l+n} \rangle = \frac{(-1)^n}{3^{|n|}},
        \quad
        S_{ZZ}(q)=\frac{4}{5+3 \cos q}.
\end{equation}

Figure~1 shows finite-$N$ data and the exact result for $N=\infty$ of each
static structure factor. At the $q$-values realized for $N=12, 15, 18$, only the
trimer data [panel (d)] are subject to finite-size corrections.

\begin{figure}[htb]
\centerline{\epsfig{file=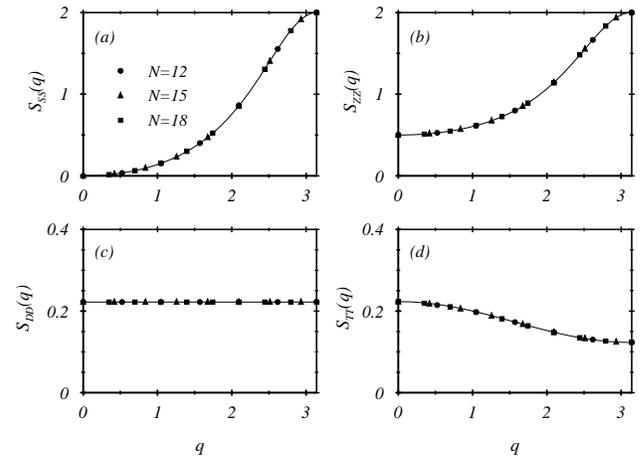,width=6.8cm,angle=-90}}
\caption[1]
{Static structure factors for the fluctuation operators (a) $F_q^S$ (spin), (b)
  $F_q^Z$ (center), (c) $F_q^D$ (dimer), (d) $F_q^T$ (trimer) in the VBS ground
  state of $H_\theta$ at $\theta_{VBS}=\arctan{1\over 3}$ for $N=12$
  ({\Large$\bullet$}), $N=15$ ({\large$\blacktriangle$}), $N=18$
  ($\blacksquare$) with periodic boundary conditions, and for $N=\infty$ (solid
  lines).}\label{fig:1}
\end{figure}

At the VBS point, no trace exists of any of the four ordering tendencies which
become important in one or the other part of the parameter space. The dimer and
trimer correlations are zero for distances $|n|\geq 1$ and $|n|\geq 2$,
respectively, while the spin and center correlations are nonzero but have a very
short correlation length $(\xi=\ln 3=1.0986\ldots)$. The topological long-range
order in the VBS state,\cite{WH93,SJG96} described by a string correlation
function, is a different matter not discussed here.
%
%
\subsection{Dynamic structure factors}\label{sec:dsf}
%
%
The topological long-range order known to be strongest in the VBS state provides
an environment, as we shall see, where point-like elementary excitations can
propagate freely and corresponding stationary states (magnons) form a branch
with well-defined dispersion. In the following, these elementary excitations and
several kinds of composite excitations made of magnon pairs will be probed from
different angles by the four fluctuation operators $F_q^A$.

In Fig.~2 we display $\omega_\lambda^A$ versus $q$ of the dynamically relevant
spin, center, dimer, and trimer excitation spectra as obtained from the
finite-size continued-fraction analysis for $N=12,15,18$.\cite{VM94,FKMW95}. The
relative spectral weight $w_\lambda^A\equiv W_\lambda^A/S_{AA}(q)$ is indicated
by the size of the data points. All four spectra are different from each other,
but the spin and center spectra [panels (a) and (b)] share some features as do
the dimer and trimer spectra [panels (c) and (d)]. The commonalities and
differences yield important clues about the composition of each spectrum.

\widetext
\begin{figure*}[htb]
\centerline{\epsfig{file=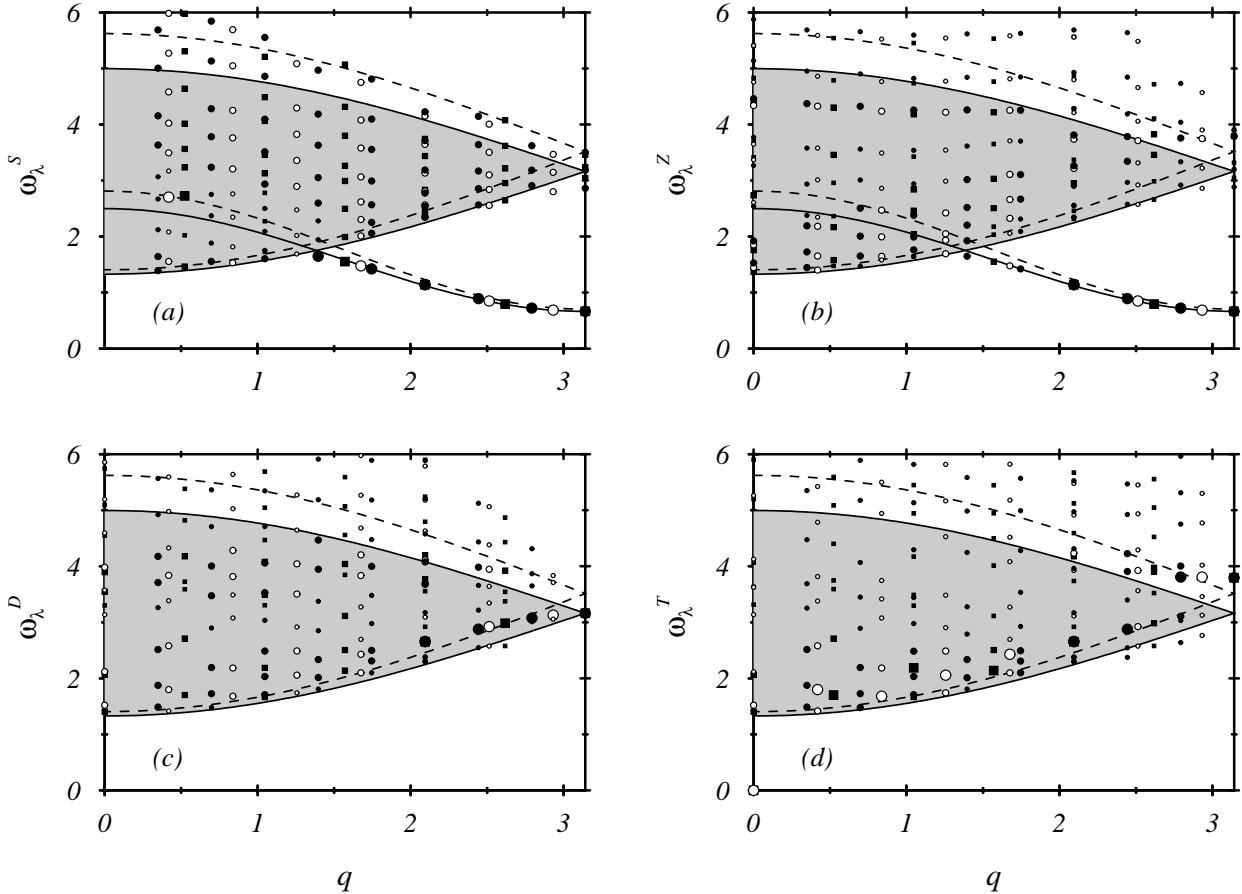,width=14cm,angle=-90}}
\caption[2]
{Frequency $\omega_\lambda^A$ versus wave number $q$ of the finite-$N$
  excitations [$N=12$ ({\Large$\bullet$}), $N=15$ ({\Large$\circ$}), $N=18$
  ($\blacksquare$)] which carry most of the spectral weight in the $T=0$ dynamic
  structure factors $S_{AA}(q,\omega)$ for (a) the spin, (b) the center, (c) the
  dimer, and (d) the trimer fluctuations of $H_\theta$ with $J=1$ at
  $\theta=\arctan{1\over 3}$. The three sizes of symbols used distinguish
  excitations with relative spectral weight in the ranges $w_\lambda^A \geq 0.5$
  (large), $0.5 > w_\lambda^A \geq 0.1$ (medium), $0.1 > w_\lambda^A \geq
  0.001$ (small). The solid and dashed lines represent one-magnon dispersions and
  two-magnon continuum boundaries as explained in the text.}\label{fig:2}
\end{figure*}
\narrowtext

The low-frequency region at $q\gtrsim \pi/2$ in the spin and center spectra is
dominated by a branch of one-magnon states, which have $S_T=1$ and, therefore, do
not make their appearance in the dimer and trimer spectra. At $q\lesssim \pi/2$
the one-magnon states overlap in energy with what will be identified as a
continuum of two-magnon states. Here the magnon interaction precludes the
observation of individual one-magnon states. Outside the region of overlap, i.e.
for $q\gtrsim \pi/2$, the one-magnon states carry more than 95\% of the spectral
weight in $S_{SS}(q,\omega)$ and $S_{ZZ}(q,\omega)$.
%
%
\subsection{Single-mode approximation}\label{sec:sma}
%
%
An exact single-mode spectrum at any given $q$-value would lead to a spontaneous
termination of the recursion algorithm in the first iteration.  Terminating it
forcibly by setting $b_1^2(q)=0$ in (\ref{d0AA}) is equivalent to invoking the
single-mode approximation. In general, this is a dubious scheme, but it may have
some merits if the spectral-weight distribution is known to be dominated by a
single mode.  In the present context, the exact calculation of the first
continued-fraction coefficient in (\ref{d0AA}) yields the following single-mode
dispersion of the VBS magnons:\cite{AAH88}
\begin{equation}\label{wMSM}
        \omega_M^{(SM)}(q) = J\frac{\sqrt{10}}{9}(5+3\cos q).
\end{equation}
This prediction, which is shown as a dashed curve in panels (a) and (b) of
Fig.~2, fits the finite-$N$ data less than perfectly. The value
$\omega_M^{(SM)}(\pi)/J = 0.70272...$, for example, exceeds the value
$\omega_M(\pi)/J = 0.66433(2)$ of the extrapolated\cite{note6} finite-size spin
excitation energy by more than 5\%. 

There are three kinds of two-magnon scattering states formed by pairs of one-magnon
triplets: states with $S_T=1$, which contribute to $S_{SS}(q,\omega)$ and
$S_{ZZ}(q,\omega)$, states with $S_T=0$, which contribute to $S_{DD}(q,\omega)$
and $S_{TT}(q,\omega)$, and states with $S_T=2$, which are observable in
$S_{ZZ}(q,\omega)$ only. The magnon interaction is different in each $S_T$
subspace. Free two-magnon states form a 2-parameter continuum
\begin{equation}\label{w2Mkq}
        \omega_{2M}(k,q) \equiv \omega_M(q/2-k) + \omega_M(q/2+k),
\end{equation}
in $(q,\omega)$-space. The continuum boundaries are determined by the solution
of $\partial\omega_{2M}/\partial k = 0$. The boundaries thus predicted by the
single-mode dispersion (\ref{wMSM}),
\begin{equation}\label{wpmSM}
  \omega_\pm^{(SM)}(q)\equiv\left.\omega_{2M}^{(SM)}(k,q)\right|_{k=0,\pi}
  =\frac{2\sqrt{10}}{9}J\left(5\pm3\cos\frac{q}{2}\right),
\end{equation}
are shown dashed in all four panels of Fig.~2. The coalescence of this 
two-magnon continuum into one spectral line at $q=\pi$ is a consequence of the
non-generic symmetry property
\begin{equation}\label{wMpwM}
        \omega_M(q) + \omega_M(\pi-q) = \text{const}
\end{equation}
of the single-mode one-magnon dispersion $\omega_M^{(SM)}(q)$.
%
%
\subsection{Improved 1$-$magnon dispersion}\label{sec:i1md}
%
%
The key to a significant improvement of the one-magnon dispersion and the
two-magnon continuum boundaries over the single-mode predictions (\ref{wMSM})
and (\ref{wpmSM}) can be found in the dimer spectrum, i.e. in Fig.~2(c). This
spectrum does indeed collapse into a single spectral line at $q=\pi$. The
finite-$N$ analysis demonstrates beyond ambiguity that there exists an exact
eigenstate, $|D\rangle=F_\pi^D|G\rangle$ with an $N$-independent excitation
energy $\omega_D=\sqrt{10}J$, which carries all the spectral weight of
$S_{DD}(\pi,\omega)$. We interpret this to be a noninteracting two-magnon
singlet.\cite{note3} Its energy is significantly lower than the single-mode
prediction, $\omega_\pm^{(SM)}(\pi)=(10/9)\sqrt{10}J$.

We now use the exact two-magnon excitation energy, $\omega_\pm(\pi)=\omega_D$,
in conjunction with the extrapolated value,
$\omega_M(\pi)=0.66433(2)J$, of the one-magnon excitation gap to construct a
modified one-magnon dispersion of the form
\begin{equation}\label{impdsp}
        \omega_M(q) = J(a + b\cos q)
\end{equation}
with $a=\sqrt{5/2}=1.58113...$ and $b= 0.91681(2)$, which still
satisfies the symmetry property (\ref{wMpwM}). This dispersion is
shown as a solid line in panels (a), (b) of Fig.~2 and the resulting two-magnon
continuum (\ref{w2Mkq}) with boundaries 
\begin{equation}
      \omega_\pm(q)=2J\left(a\pm b\cos{q\over 2}\right)
\end{equation}
as a shaded region in panels (a)-(d). The finite-$N$ data for the one-magnon
states are fitted almost perfectly by (\ref{impdsp}). The energies of all $N=18$
states that can be clearly identified as one-magnon excitations are listed in
Table~I. The relative deviation of the single-mode prediction
(\ref{wMSM}) and the improved one-magnon dispersion (\ref{impdsp}) from these
finite-$N$ data are listed for comparison.

\begin{table}[htb]
 \caption{One-magnon excitation energies of $H_\theta$ at
   $\theta=\arctan\frac{1}{3}$ for wave numbers $q=(2\pi/N)\lambda$ and
   relative deviation of the predictions (\ref{wMSM}) and
   (\ref{impdsp}).}\label{T1} 
  \begin{tabular}{cccc}
    $\lambda$ & $\omega_\lambda/J$ & $\omega_M^{(SM)}(q)/\omega_\lambda -1$ &
    $\omega_M(q)/\omega_\lambda -1$ \\ \hline
    3 & 2.092(2)   & 0.091 & $-0.025$ \\
    4 & 1.649(2)   & 0.176 & $+0.055$ \\
    5 & 1.4235(2)  & 0.106 & $-0.001$ \\
    6 & 1.14009(2) & 0.079 & $-0.015$ \\
    7 & 0.89071(2) & 0.066 & $-0.013$ \\
    8 & 0.72314(2) & 0.060 & $-0.005$ \\
    9 & 0.66433(2) & 0.058 & used to fit (\ref{impdsp}) 
  \end{tabular}
\end{table}

The lower two-magnon boundary $\omega_-(q)$ is now in much better agreement with
the finite-$N$ two-magnon spectral thresholds in the invariant subspaces with
$S_T=0,1,2$. Only for $q\lesssim\pi$ do we observe finite-$N$ excitations with
significant spectral weight which stray below the predicted two-magnon region in
all four panels. Here the spectral threshold of the three-magnon continuum as
inferred from (\ref{impdsp}) is lower than the two-magnon continuum, with a
minimum energy $3\omega_M(\pi)/J\simeq 2.0$ at $q=\pi$. However, since the stray
states are located near the lower two-magnon boundary, it is more likely that
they are interacting two-magnon states than three-magnon states. The nonzero
two-magnon bandwidth at $q=\pi$ in the $S_T=1$ subspace as observed in the
finite-$N$ data of panels (a) and (b) is perhaps the most compelling evidence of
the magnon interaction in this region of $(q,\omega)$-space.

A curious phenomenon occurs in the $S_T=0$ spectrum of the trimer fluctuations
as shown in panel (d). At $q\neq\pi$ all dynamically relevant excitations are
identical to those observed in the dimer fluctuations, understandably with
different matrix elements. At $q=\pi$, the trimer spectrum collapses into a
single spectral line as does the dimer spectrum, but the trimer line has a
higher energy, $\omega_T/J=12/\sqrt{10}$, which is again $N$-independent. Here
we have another exact eigenstate $|T\rangle = F_\pi^T|G\rangle$. Since
$|T\rangle$ is necessarily orthogonal to $|D\rangle$, the former state cannot
contribute to the dimer fluctuations, $\langle G|F_\pi^D|T\rangle=0$, and the
latter not to the trimer fluctuations, $\langle G|F_\pi^T|D\rangle=0$.

The energy of the state $|T\rangle$ is evidently beyond the range of the
two-magnon continuum. If it is formed by two magnons, then it must be a bound
state with positive interaction energy.\cite{note7} The data shown in panel (d)
indicate that the dominant $S_T=2$ contribution to $S_{ZZ}(\pi,\omega)$ comes
from a state that is degenerate with the lone trimer excitation $|T\rangle$.
%
%
\subsection{Two$-$magnon scattering states and bound states}\label{sec:tmsb}
%
%
In their DMRG study of selected excited states of $H_\theta$ for $\theta=0$,
White and Huse\cite{WH93} observed that at $q$ near $\pi$ the magnon interaction
is attractive for two-magnon states with $S_T=1$ and repulsive for two-magnon
states with $S_T=0$ and $S_T=2$. The finite-$N$ data for
$\theta=\arctan\frac{1}{3}$ in Fig.~2 exhibit trends that are similar in some
respects yet different in others.

Near the bottom of the two-magnon region, the interaction is found to be very
weak in all three $S_T$ subspaces, and neither uniformly attractive nor
uniformly repulsive in any $S_T$ subspace. At higher energies, the interaction
is stronger and repulsive in all three subspaces. The observation that the
dynamically relevant excitations in panels (b), (c), (d) spread to higher
energies than those in panel (a) indicates that the (positive) interaction
energy is considerably larger for $S_T=0, 2$ than for $S_T=1$.

Several excitations, mainly near the bottom of the shaded areas in
Fig.~\ref{fig:2}, can be identified as almost free two-magnon states. There are
28 two-magnon combinations of the one-magnon states listed in
Table~\ref{T1}. All of them can be found with appreciable spectral weight in at
least one of the four excitation spectra within a 5\% margin of interaction
energy. In Table~\ref{T2} we have listed the excitation energies and interaction
energies of these states.  Inevitably, some of the assignments made are
ambiguous. The uncertainty in any of the excitation energies listed is estimated
to be under 0.5\%.

\begin{table}[htb]
  \caption{Selected excitations with energy $\omega_{\lambda_1\lambda_2}^A$
    and wave number $q$ in the spin, center, dimer, and trimer spectra of
    $H_\theta$ with $J=1$, $\theta=\arctan\frac{1}{3}$ for $N=18$ in comparison
    with the corresponding free two-magnon combinations at energy
    $\omega_{\lambda_1}+\omega_{\lambda_2}$ and the same wave number.}\label{T2} 
  \begin{tabular}{ccc|l|l|l} 
$\lambda_1\lambda_2$ & $\omega_{\lambda_1}\!+\!\omega_{\lambda_2}$ & $qN/2\pi$ &
$\omega_{\lambda_1\lambda_2}^S$ & $\omega_{\lambda_1\lambda_2}^Z$ &
$\omega_{\lambda_1\lambda_2}^{D,T}$  \\ \hline
    99 &1.329& 0 &     & 1.35&      \\
    89 &1.387& 1 & 1.39&     &      \\
    88 &1.446& 2 &     & 1.47& 1.48 \\
    79 &1.555& 2 & 1.55&     &      \\
    69 &1.804& 3 & 1.75& 1.76&      \\
    78 &1.614& 3 & 1.60& 1.64&      \\
    59 &2.088& 4 &     &     & 2.01 \\
    68 &1.863& 4 & 1.94& 1.92&      \\
    77 &1.781& 4 &     &     & 1.81 \\
    49 &2.314& 5 &     & 2.31& 2.31 \\
    58 &2.147& 5 &     & 2.09& 2.10 \\
    67 &2.031& 5 & 2.06&     &      \\
    39 &2.757& 6 & 2.78&     &      \\
    48 &2.373& 6 &     &     & 2.37 \\
    57 &2.314& 6 & 2.35& 2.35&      \\
    66 &2.280& 6 &     &     & 2.30 \\
    33 &4.185& 6 & 4.15&     &      \\
    38 &2.816& 7 & 2.85&     &      \\
    47 &2.540& 7 &     &     & 2.55 \\
    56 &3.073& 7 &     & 2.97&      \\
    34 &3.742& 7 &     & 3.78&      \\
    37 &2.983& 8 &     &     & 3.02 \\
    46 &2.790& 8 &     &     & 2.65 \\
    55 &2.847& 8 & 2.91& 2.92&      \\
    35 &3.516& 8 & 3.62&     &      \\
    44 &3.299& 8 &     & 3.33&      \\
    36 &3.233& 9 & 3.19& 3.20&      \\
    45 &2.564& 9 & 2.44& 2.44&     
  \end{tabular}
\end{table}

The number of states belonging to a 2-parameter continuum is of O($N^2$), i.e.
of O($N$) for fixed $q$. If the integrated intensity $S_{AA}(q)$ is finite and
nonzero for any particular $q$, and if the continuum is to carry a nonzero
fraction of the spectral weight in $S_{AA}(q,\omega)$, then the average
transition rate of the continuum states must scale like $N^{-1}$. Whereas this
particular scaling behavior cannot be verified reliably on the basis of the
available data, we have been able to track several of the nearly free two-magnon
states at $q=2\pi/3$ over the four system sizes $N=9, 12, 15, 18$ and to show
that the transition rate tends to converge to zero.

The excitation energy and the relative spectral weight of one such state as
observed in the dynamic dimer and trimer structure factors are listed in
Table~\ref{T3}.\cite{note2}

\begin{table}[htb]
 \caption{Excitation energy and relative spectral weight of one finite-$N$
   two-magnon continuum state as observed in $S_{DD}(2\pi/3,\omega)$ and
   $S_{TT}(2\pi/3,\omega)$.}\label{T3}
  \begin{tabular}{r|cc|cc}
    $N$ & $\omega_\lambda^D$ & $w_\lambda^D$ 
        & $\omega_\lambda^T$ & $w_\lambda^T$ 
     \\ \hline
     9 & 2.34 & 0.112 & 2.34 & 0.249 \\
    12 & 2.31 & 0.042 & 2.31 & 0.095 \\
    15 & 2.30 & 0.020 & 2.30 & 0.046 \\
    18 & 2.30 & 0.015 & 2.30 & 0.031 \\ 
  \end{tabular}
\end{table}

The strong repulsive magnon interaction for $S_T=0$ and $S_T=2$ as indicated by
our data, raises the possibility that discrete branches of bound two-magnon
states split off the top of the two-magnon continuum in these two subspaces.

Comparing panels (a) and (b) of Fig.~\ref{fig:2} at frequencies
$3\lesssim\omega/J\lesssim 5$, we see that the dynamically relevant finite-$N$
excitations are arranged in contrasting patterns. In panel (a) we have an
arrangement of points which is typical of a 2-parameter continuum. As $N$
increases, more points are added and spread roughly evenly along the frequency
axis. In panel (b), by contrast, the data points are arranged in branches with
an almost $N$-independent separation.

The number of states belonging to a discrete branch is of O($N$), i.e. one state
per wave number over the range of the branch. The transition rate of such a
state will converge to a nonzero value as $N\rightarrow\infty$ for any branch
that carries a nonzero fraction of the integrated intensity. This convergence
can be demonstrated most convincingly for the one-magnon state at $q=2\pi/3$,
because it is far removed in energy from any other state in the same invariant
subspace. Our data for $N=9, 12, 15, 18$ yield $\omega_\lambda/J\rightarrow
1.14009(1)$, $w_\lambda^S\rightarrow 0.95124(1)$, $w_\lambda^Z\rightarrow
0.53507(2)$. 

Several higher lying states in $S_{ZZ}(2\pi/3,\omega)$ which appear to be part
of discrete branches have been tracked over the system sizes $N=9, 12, 15, 18$.
Their transition rates also tend to converge to a nonzero value. The data for
two such states are compiled in Table~\ref{T4}.

\begin{table}[htb]
 \caption{Excitation energy and spectral weight as observed in
    $S_{ZZ}(2\pi/3,\omega)$, $S_{DD}(2\pi/3,\omega)$, or $S_{TT}(2\pi/3,\omega)$
    of several finite-$N$ excited states that seem to belong to discrete
    branches in the limit $N\rightarrow\infty.$}\label{T4}
  \begin{tabular}{r|cc|cc|cc}
    $N$ & $\omega_\lambda^Z$ & $w_\lambda^Z$ 
        & $\omega_\lambda^D$ & $w_\lambda^D$ 
        & $\omega_\lambda^T$ & $w_\lambda^T$  \\ \hline
     9  & 3.259 & 0.050 & 2.723 & 0.486 & 2.723 & 0.475 \\
     12 & 3.369 & 0.082 & 2.654 & 0.540 & 2.654 & 0.611 \\
     15 & 3.215 & 0.112 & 2.654 & 0.525 & 2.654 & 0.595 \\
     18 & 3.265 & 0.117 & 2.661 & 0.532 & 2.660 & 0.594 \\ \hline
     9  & 3.571 & 0.151 & 4.317 & 0.150 & 4.317 & 0.181 \\
     12 & 3.786 & 0.108 & 4.196 & 0.160 & 4.179 & 0.132 \\
     15 & 3.736 & 0.110 & 4.114 & 0.185 & 4.233 & 0.108 \\
     18 & 3.812 & 0.114 & 4.078 & 0.193 & 4.313 & 0.091 
  \end{tabular}
\end{table}

The arrangement of points in panels (c) and (d) of Fig.~\ref{fig:2} looks more
irregular. Evidence for the layered structure typical of discrete branches can
be discerned at high frequencies $(\omega/J\gtrsim5)$. At lower frequencies,
some of the dynamically relevant states have already been identified as nearly
free two-magnon continuum states. But then we can also observe states with a
fairly large spectral weight whose $N$-dependence indicates that they belong to
a discrete branch.

The most prominent such branch in panel (c) has its endpoint at $q=\pi$ in the
exact dimer excitation $|D\rangle$ discussed previously. A corresponding branch
which ends in the exact trimer excitation $|T\rangle$ at $q=\pi$ can be
discerned in panel (d). 

It may well be the case that the wave function of the eigenstate
$|T\rangle=F^T_\pi|G\rangle$ has bound-state character with positive interaction
energy and the eigenstate $|D\rangle=F^D_\pi|G\rangle$ scattering-state character
(with zero interaction energy). A clear-cut distinction between the wave
functions of bound states and scattering states is known to exist even for
finite $N$ in the $s=1/2$ Heisenberg ferromagnet, for example, and can be
described in the framework of the Bethe ansatz.\cite{KM97}

%
%
\section{Haldane phase}\label{sec:hal}
%
%
%
%
\subsection{Static structure factors}\label{sec:sssf}
\begin{figure*}[ht]
  \centerline{\epsfig{file=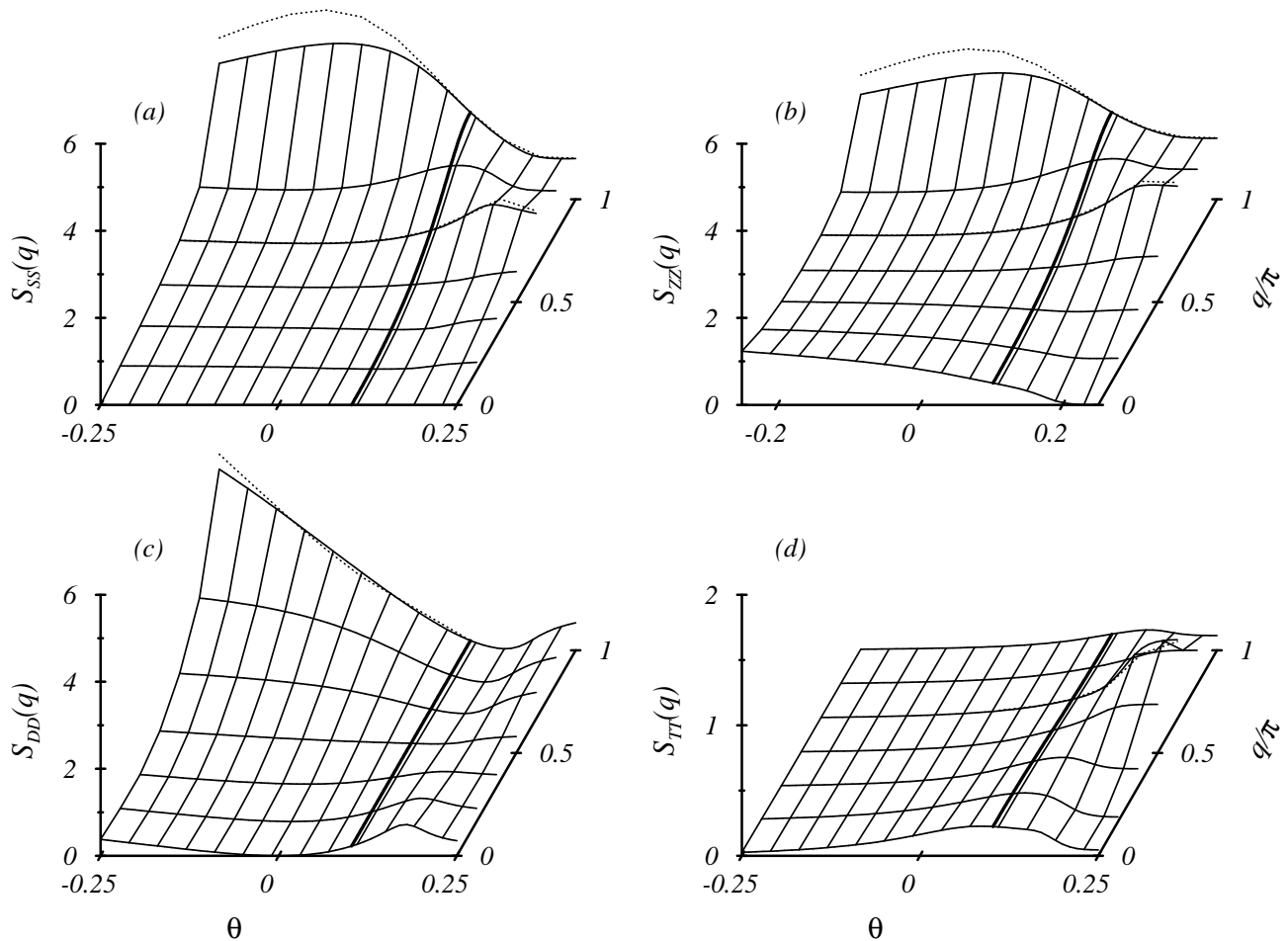,width=14cm,angle=-90}}
  \caption{Static structure factors for the fluctuation operators (a) $F_q^S$
    (spin), (b) $F_q^Z$ (center), (c) $F_q^D$ (dimer), (d) $F_q^T$ (trimer)
    plotted versus $q$ and $\theta$ for $N=12$ over the range
    $-\pi/4\leq\theta\leq\pi/4$. The dotted lines represent $N=18$ data for
    $q=2\pi/3, \pi$.}\label{fig:4}
\end{figure*}

%
%
Any change of the Hamiltonian parameter $\theta$ away from the value
$\theta_{VBS}$ in the interior of the Haldane phase toward lower or higher
values softens the topological long-range order and thus enhances specific
quantum fluctuations in the ground state. The contrasting enhancements of
fluctuations for decreasing or increasing $\theta$-values, which reflect the
different ordering tendencies near the predicted phase boundaries at
$\theta=-\pi/4$ and $\theta\lesssim\pi/4$, respectively, can be observed to a
certain extent in the finite-$N$ static structure factors for the spin, center,
dimer, and trimer correlations, as shown in Fig.~\ref{fig:4}. The finite-$N$
effects are minute except for $q=2\pi/3$ at $\theta \simeq \pi/4$ and for
$q=\pi$ at $\theta < 0$.
\widetext
\narrowtext

In the long-wavelength limit $q\rightarrow 0$, the function $S_{SS}(q)$ is
observed to vanish for all $\theta$-values covered in Fig.~\ref{fig:4}, whereas
$S_{DD}(q)$ vanishes only at $\theta=0$, $S_{ZZ}(q)$ only at $\theta=-\pi/4$,
and $S_{TT}(q)$ not at all. These properties are the result of different
conservation laws: 

The $z$-component of the total spin, $S_T^z\equiv\sum_l S_l^z$, is a constant of
the motion for all $\theta$-values, and it has eigenvalue zero in the singlet
ground state. At $\theta=0$, the operator $D_T\equiv\sum_l D_l$ with $D_l$ as
defined in (\ref{Dl}) can be written as $D_T=(H-E_0/J)$ and is, therefore, a
constant of the motion with zero eigenvalue.

The operator $Z_T\equiv\sum_l Z_l$ with $Z_l$ as defined in (\ref{Zl})
commutes with $H_{\pi/4}$. Furthermore, the Bethe ansatz demonstrates that all
eigenvectors of $H_{\pi/4}$ have definite numbers $n_1$, $n_2$, $n_3$ of up,
zero, and down spins, respectively, on the $N$ sites of the lattice. The ground
state has $n_1=n_2=n_3=N/3$, which implies that $Z_T$ has eigenvalue zero.

As $\theta$ decreases from the value $\theta_{\small VBS}$, the spin and dimer
fluctuations at $q=\pi$ experience a gradual enhancement, which reflects a
continuous increase in the spin and dimer correlation lengths.\cite{note10} 
On approach of the phase boundary at $\theta=-\pi/4$, these correlation lengths
diverge. At $\theta=-\pi/4$ the spectra with $S_T=0, 1$ are gapless, and the
static spin and dimer structure factors exhibit cusp singularities at $q=\pi$.

At this critical point, the spin and dimer ordering tendencies are in
competition. A perturbation of $H_{-\pi/4}$ can produce N\'eel long-range order
or dimer long-range order. Uniaxial exchange anisotropy, for example, is likely
to produce N\'eel long-range order, whereas the further strengthening of the
biquadratic exchange $(\theta\lesssim-\pi/4)$ is almost certain to produce dimer
long-range order.\cite{FS95}

In the regime $-\pi/4\leq\theta\leq\theta_{\small VBS}$ of the Haldane phase,
the center structure factor exhibits properties very similar to those of the
spin and dimer structure factors. It remains to be seen whether the $q=\pi$
enhancement in the center structure factor is due primarily due to the $S_T=1$
excitations, which also dominate the spin structure factor or whether the
$S_T=2$ spectrum contributes significantly to the spectral intensity. The trimer
structure factor, by contrast, remains featureless throughout this regime. The
conclusion is that there are no significant trimer fluctuations in this part of
the phase diagram.

In the regime $\theta_{\small VBS}\leq\theta\leq\pi/4$, it is the dimer
fluctuations that remain insignificant and the dimer structure factor that
remains featureless. Here the prevailing ordering tendencies are captured by the
spin, center and trimer structure factors. In all three of them a distinct
enhancement in intensity at $q=2\pi/3$ emerges as $\theta$ increases toward the
Lai-Sutherland point $\theta=\pi/4$.

The DMRG study of Schollw\"ock et al.\cite{SJG96} showed that as $\theta$
increases from $\theta_{\small VBS}$ the spin correlations start to become
incommensurate and that the correlation length increases gradually. The
underlying periodicity moves gradually from $q=\pi$ at $\theta_{\small VBS}$ to
$q=2\pi/3$ at or before $\theta_{LS}$. Because of the short-range nature of the
spin correlations, the change of periodicity in real space is not fully
synchronized with the movement of the peak of the static structure factor in
reciprocal space.

The phase at $\theta>\pi/4$ has been named ``trimerized'' phase with no
compelling evidence in support of trimer ordering, mainly because the dominant
fluctuations have period three. However, there is evidence that the dominant
fluctuations at $q=2\pi/3$ are not those that are probed by the trimer
fluctuation operator.

One piece of evidence is provided by the $N$-dependence of the static spin,
trimer and center structure factors in the vicinity of the phase boundary at
$\theta\lesssim\pi/4$, whose exact position is not firmly established.  These
values are listed in Table~\ref{T5}. The trimer intensities extrapolate
downward, whereas the spin and center intensities extrapolate upward. The
highest intensity is observed in the center structure factor.\cite{note11}

\begin{table}[ht]
 \caption{Static spin, trimer, and center structure factors at $q=2\pi/3$ for 
   finite $N$ and extrapolated values.\cite{note6}}\label{T5}
  \begin{tabular}{r|cc|cc|cc}
    $N$ & \multicolumn{2}{c|}{$S_{SS}(2\pi/3)$} & 
          \multicolumn{2}{c|}{$S_{TT}(2\pi/3)$} & 
          \multicolumn{2}{c}{$S_{ZZ}(2\pi/3)$} \\ \hline
    $\theta/\pi$ & 0.20 & 0.25 & 0.20 & 0.25 & 0.20 & 0.25 \\ \hline
      9  & 1.3386 & 1.2068 & 0.5803 &  0.6288    & 1.8166 & 1.8102 \\
      12 & 1.4366 & 1.2507 & 0.5390 &  0.6049    & 1.8899 & 1.8761 \\
      15 & 1.5145 & 1.2825 & 0.5151 &  0.5952    & 1.9440 & 1.9237 \\
      18 & 1.5778 & 1.3070 & 0.4984 &  0.5906    & 1.9854 & 1.9606 \\ \hline
$\infty$ & 2.15(3)& 1.76(3)& 0.44(2)&  0.58(2)   & 2.28(1)& 2.64(5)
  \end{tabular}
\end{table}

Further evidence which speaks against the trimer nature of the phase at
$\theta>\pi/4$ can be inferred from the finite-$N$ excitation spectrum at
$q=2\pi/3$, where the trimer order parameter fluctuations are expected to be
come dominant. It turns out that in this parameter regime the lowest excitation
at $q=2\pi/3$ is not a state with $S_T=0$, which could contribute to
$S_{TT}(q,\omega)$, nor is it a state with $S_T=1$, which could contribute to
$S_{SS}(q,\omega)$. From Ref. \onlinecite{FS91} we know that it is a state with
  $S_T=2$, and our dynamics data show that this excitation contributes to
  $S_{ZZ}(q,\omega)$ with a large transition rate.

%
%
\subsection{One-magnon and two-magnon states}\label{sec:omtms}
%
%
The finite-$N$ data for the dynamic spin, center, dimer, and trimer structure
factors throughout the Haldane phase indicate that we must distinguish two
regimes. For $\theta_{VBS}\leq\theta\leq\pi/4$, the structure of the low-lying
excitations as seen through the lenses of all four dynamic structure factors
undergoes a major metamorphosis. The analysis of this spectrum and its relation
to the prevailing ordering tendency will be the focus of a separate study that
uses the exactly solvable case $\theta=\pi/4$ as the starting point.\cite{note8}

Over much of the range $-\pi/4 < \theta < \theta_{VBS}$, the dynamically
relevant spectrum continues to be dominated by one-magnon and two-magnon states
with properties that connect seamlessly to the results for $\theta_{VBS}$
established in Sec.~\ref{sec:vbs}. However, at $\theta < \theta_{VBS}$, the
two-parameter ansatz (\ref{impdsp}) for the one-magnon dispersion is now
inadequate because the symmetry (\ref{wMpwM}) is no longer supported by the
finite-$N$ data. The absence of this symmetry at $\theta < \theta_{VBS}$ is also
indicated by the manifestly nonzero width at $q=\pi$ of the two-magnon continua.

We propose to use instead the 3-parameter ansatz
\begin{equation}
\omega_M(q)=J\sqrt{(a+b_1\cos q)(a+b_2\cos q)}
\label{tpa}
\end{equation}
with $a>0$ and $|b_1|,|b_2|\leq a$ as an interpolation formula between
$\theta=\theta_{VBS}$, where we have $a=\sqrt{5/2}$, $b_1=b_2 \equiv b =
0.91681(2)$ as discussed previously, and $\theta=-\pi/4$, where (\ref{tpa}) with
$a=b_1=-b_2=\pi$ describes magnon states having turned into spinon states that
are amenable to a rigorous analysis in the framework of the Bethe
ansatz.\cite{Takh82}

Interestingly, the 3-parameter dispersion is already known to have at least two
exact realizations in the dynamics of spin chains. (i) The magnon dispersion of
the classical spin-$s$ $XYZ$ antiferromagnet with $J_x \geq J_y \geq J_z$ is
given by\cite{KTM82}
\begin{equation}\label{clsw}
\omega_{XYZ}^{CL}(q) = 2s\sqrt{(J_x - J_y\cos q)(J_x + J_z\cos q)}.
\end{equation}
(ii) The one-particle dispersion in the fermion representation of the $s=1/2$
$XY$ model with couplings $J_\pm = \frac{1}{2}(J_x\pm J_y)$ and magnetic field
$h_z$ is given by\cite{LSM61}
\begin{equation}\label{qufr}
\omega_{XY}^{QU}(q) = \sqrt{(h_z + J_+\cos q)^2 + J_-^{2}\sin^2q},
\end{equation}
which is, for restricted $h_z,J_\pm$, equivalent to (\ref{tpa}). In the $XY$
model, the special case $h_z^2=J_+^2-J_-^2$, where (\ref{qufr}) becomes a linear
function of $\cos q$, has been identified as a disorder point, where the spin
fluctuations are minimally correlated,\cite{KTM82,TM83} such as is also the case
at the VBS point of the spin-1 chain (\ref{H}).

The deformation of the one-magnon dispersion over the range $-\pi/4 < \theta <
\theta_{VBS}$ has a strong effect on the shape of the associated two-magnon
continuum. Of special interest are the excitation gaps at $q=0$ and
$q=\pi$. Their dependence on the parameters $a,b_1,b_2$ is
\begin{mathletters}\label{gap3p}
\begin{eqnarray}
\omega_\pm(0) &=& J\sqrt{(a \pm b_1)(a \pm b_2)}, \\
\omega_+(\pi) &=& 2Ja, \\
\omega_-(\pi) &=& 
J\left[\sqrt{(a+b_1)(a+b_2)} \right. \nonumber \\
&& ~~\left.+ \sqrt{(a-b_1)(a-b_2)} \right].
\end{eqnarray}
\end{mathletters}

The adequacy of the 3-parameter dispersion for the description of the
dynamically dominant excitation spectrum must be judged on the two requirements
(i) that it yields a reasonable fit of the one-magnon energies and (ii) that the
associated two-magnon continuum covers the range of the corresponding
finite-chain data, especially near the spectral threshold.

\begin{figure*}[htb]
\centerline{\epsfig{file=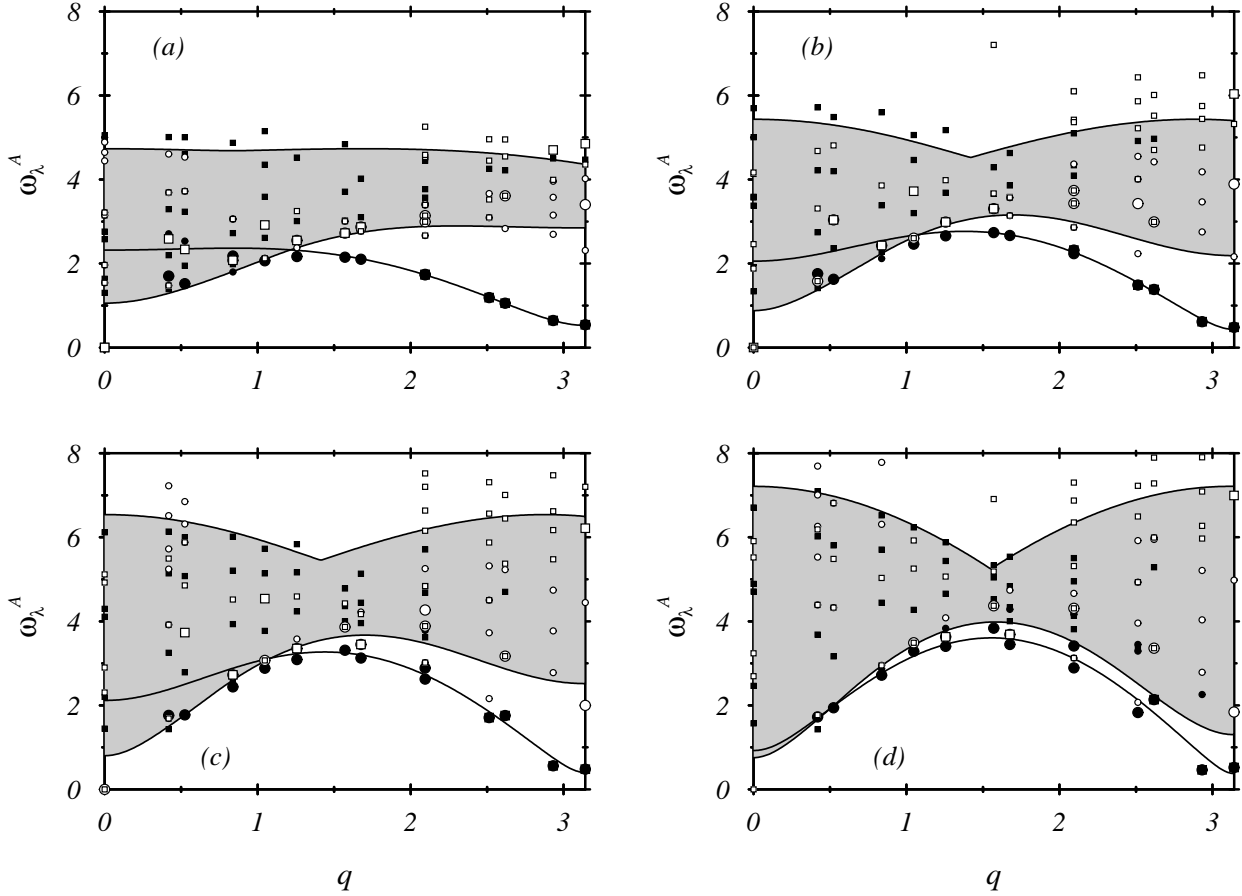,width=14cm,angle=-90}}
\caption[5]
{Frequency $\omega_\lambda^A$ versus wave number $q$ of the finite-$N$
  excitations [$N=12$ and $N=15$] which carry most of the spectral weight in the
  $T=0$ dynamic structure factors $S_{AA}(q,\omega)$ for the spin
  ({\Large$\bullet$}), center ($\blacksquare$), dimer ({\Large$\circ$}), and
  trimer ($\square$) fluctuations of $H_\theta$ with $J=1$ at (a) $\theta/\pi =
  0.05$, (b) $\theta/\pi = 0$, (c) $\theta/\pi = -0.05$, (d) $ \theta/\pi =
  -0.10$. The symbol sizes are explained in the caption of Fig.~\ref{fig:2}. The
  solid lines represent the one-magnon dispersion (\ref{tpa}) and the associated
  two-magnon boundaries with parameter values (a) $a=2.18, b_1=2.09, b_2=-0.92$,
  (b) $a=2.70, b_1=2.66, b_2=-2.13$, (c) $a=3.25, b_1=3.22, b_2=-2.55$, (d)
  $a=3.61, b_1=3.58, b_2=-3.49$.}\label{fig:5}
\end{figure*}

The finite-$N$ data shown in Fig.~\ref{fig:5} for four values of $\theta$
between $\theta_{VBS}$ and $-\pi/4$ confirm that the excitations dominating
the $T=0$ dynamical properties in this regime remain well described in terms of
a branch one-magnon states and a continuum of two-magnon scattering states.
\widetext
\narrowtext

%
\acknowledgements
%
The work at URI was supported by NSF Grant DMR-93-12252 and by the Max Kade
Foundation. 

%
%
\begin{appendix}
\section{States with maximum \\ N\'eel, dimer, trimer, or center \\ 
         long range order}
\label{lro}
%
%
The order parameters associated with the four fluctuation operators $F_q^S$
(spin), $F_q^D$ (dimer), $F_q^T$ (trimer), $F_q^Z$ (center) introduced in
Sec.~\ref{sec:flucop} can be written in the form
\begin{equation} 
P_A = \frac{1}{N}\sum_{l=1}^N e^{iq_Al}A_l,
\end{equation}
where $q_S=q_D=\pi$, and $q_T=q_Z=2\pi/3$. Their exact nature is best
illustrated by those eigenvectors of each operator $P_A$ with an
eigenvalue of maximum magnitude. Every such vector $|\Phi_k^A\rangle$,
$k=1,2,\ldots$ represents the long-range order associated with order parameter
$P_A$ in its purest form. For each operator $P_A$ there must be at least two
such vectors for it to qualify as an order parameter. 

(i) There are two {\it N\'eel} states with eigenvalues $\exp(iq_{S} k)$:
\begin{equation}\label{nstates}
  |\Phi_1^S\rangle = |+-+-\cdots\rangle, \;\;\; 
  |\Phi_2^S\rangle = |-+-+\cdots\rangle.
\end{equation}

(ii) There are two {\it dimer} states with eigenvalues $\exp(iq_{D} k)/2$ :
\begin{equation}\label{dstates}
  |\Phi_1^D\rangle = |[12][34]\cdots\rangle, \;\;\;
  |\Phi_2^D\rangle = |[23][45]\cdots\rangle,
\end{equation}
where $|[l,l+1]\rangle = (|+-\rangle +|-+\rangle -|00\rangle)/\sqrt{3}$
represents a singlet state formed by two spins 1 on neighboring sites.

(iii) There are three {\it trimer} states with eigenvalues  $\exp(iq_{T} k)/3$:
\begin{eqnarray}\label{eq:trimer-states}
  |\Phi_1^T\rangle &=& |[123][456]\cdots\rangle, \;\;\;
  |\Phi_2^T\rangle = |[234][567]\cdots\rangle, \nonumber \\
  |\Phi_3^T\rangle &=& |[345][678]\cdots\rangle, 
\end{eqnarray}
where $|[l,l+1,l+2]\rangle$ as given in (\ref{lts}) represents a singlet state
formed by three spins 1 on consecutive sites.

(iv) There are six {\it center} states with eigenvalues  $\exp(iq_{Z} k)$:
\begin{eqnarray}
  |\Phi_{1}^Z\rangle &=& |0-+ \cdots\rangle, \;\;\;
  |\Phi_{2}^Z\rangle =   |+0- \cdots\rangle, \nonumber \\
  |\Phi_{3}^Z\rangle &=& |-+0 \cdots\rangle.
\end{eqnarray}

All these vectors represent non-stationary symmetry-breaking states in the
context of the Hamiltonian (\ref{H}). None of the order parameters $P_A$
commutes with $H_\theta$. The four types of long-range order are also reflected
in the orthonormal linear combinations $|A_k\rangle$ of the vectors
$|\Phi_k^A\rangle$ which restore the translational symmetry. For
$N\rightarrow\infty$ they have the form\cite{note4}
\begin{eqnarray}
|S_k\rangle &=& \frac{1}{\sqrt{2}}(|\Phi_1^S\rangle 
             + e^{iq_Sk}|\Phi_2^S\rangle), \\
|D_k\rangle &=& \frac{1}{\sqrt{2}}(|\Phi_1^D\rangle 
             + e^{iq_Dk}|\Phi_2^D\rangle), \\
|T_k\rangle &=& \frac{1}{\sqrt{3}}(|\Phi_1^T\rangle + e^{iq_Tk}|\Phi_2^T\rangle
            + e^{2iq_Tk}|\Phi_3^T\rangle), \\
|Z_k\rangle &=& \frac{1}{\sqrt{3}}(|\Phi_{1}^Z\rangle 
            + e^{iq_Zk}|\Phi_{2}^Z\rangle
            + e^{2iq_Zk}|\Phi_{3}^Z\rangle).
\end{eqnarray}
The correlation functions
\begin{equation}
  C^{A_kA_k}_{AA}(n) \equiv \langle A_k | A_{l} A_{l+n} | A_k \rangle -
        \langle A_k | A_{l} | A_k \rangle \langle A_k | A_{l+n} | A_k \rangle
\label{caan}
\end{equation}  
in the symmetry-restored eigenvectors of the order parameters are found to
be independent of $k$ and exhibit the characteristic asymptotic behavior
\begin{equation}
  C^{A_kA_k}_{AA}(n) \stackrel{n\to\infty}{\longrightarrow} \cos(q_A n) O_A^2,
\label{caasym}
\end{equation}  
if the state $A_k$ corresponds to the operator $A$. Here $O_A$ is the saturation
value of the order parameter $P_A$: $O_S^2=1$, $O_D^2=1$, $O_T=2(2/3)^6$,
$O_Z=1$. In all cases it turns out that the asymptotic behavior is reached
already at small distances $(|n|\geq 3)$.

If the state $A_k$ does not correspond to the operator $A$ in (\ref{caan}), such
as in $C_{DD}^{S_kS_k}(n)$, then these correlation functions are found to vanish
for all pairs of spins with $|n|\geq 3$. There are two exceptions to that
rule:
\begin{eqnarray}
C_{SS}^{Z_kZ_k}(n) &=& \frac{2}{3}\cos(q_Zn) \\
C_{ZZ}^{S_kS_k}(n) &=& \frac{3}{4}\cos(q_Sn)
\end{eqnarray}
This anomaly tells us that the spin fluctuation operator $F_q^S$ does not only
probe N\'eel long-range order (at $q=\pi$) but also center long-range order (at
$q=2\pi/3$) as defined in this paper. Likewise, the center fluctuation operator
$F_q^Z$, designed here to probe a particular type of fluctuation and potential
ordering, does not only see center long-range order (at $q=2\pi/3$) when such
order exists but also N\'eel long-range order (at $q=\pi$) when that is
present.

The main purpose of introducing the center fluctuation operator, which couples
to the $S_T=1$ and $S_T=2$ excitation spectra, has been to illuminate aspects of
the predominant fluctuations in the parameter range $\pi/4 < \theta < \pi/2$,
which are invisible to any of the other three fluctuation operators, which all
couple either to the $S_T=1$ spectrum (spin) or to the $S_T=0$ spectrum (dimer,
trimer). This distinctive property of $F_q^Z$ is of vital importance in view of
the fact that the lowest lying finite-$N$ excitations in this parameter range
have total spin $S_T=2$. It may well be the case that a different fluctuation
operator, which does not see the N\'eel long-range order and whose
characteristic long-range order is not picked up by the spin fluctuation
operator provides a better description of the predominant fluctuations in this
regime. However, a suitable candidate has yet to be found.

Nontrivial realizations of the fully ordered ground states are known at least
three of the four order parameters: (i) The N\'eel states (\ref{nstates}) are
stabilized in the 1D $s=1/2$ $XYZ$ model with $0<J_x=-J_y<J_z$.\cite{BCM84} (ii)
The $s=1$ dimer states~(\ref{dstates}) have $s=1/2$ counterparts, which are
constructed from corresponding singlet states: $|[l,l\!+\!1]\rangle\!=\!
(|\!\uparrow\downarrow\rangle\!-\!|\!\downarrow\uparrow\rangle)/\sqrt{2}$. The
dimer ground-states are realized in the $s=1/2$
Hamiltonian\cite{MG69,AKLT88}
\begin{equation}\label{eq:HMG}
H =\sum_{n=1}^{N} \left({\cal P}_{n,n+1}+\frac{1}{2}{\cal P}_{n,n+2}\right),
\end{equation}
with nearest and next-nearest neighbor interactions, here expressed in terms of
permutation operators
\begin{equation}\label{eq:Pnm-12}
{\cal P}_{n,m} = \frac{1}{2} +2 {\bf S}_n \cdot {\bf S}_m.
\end{equation}
(iii) It can be shown that the trimer states (\ref{eq:trimer-states}) are
eigenstates of the Hamiltonian
\begin{eqnarray}\label{eq:HT}
    H_T &=& \sum_{n=1}^{N} \left(
      {\cal P}_{n,n+1}+\frac{1}{2}{\cal P}_{n,n+2}+\frac{1}{4}{\cal P}_{n,n+3}
      \nonumber \right.\\ && 
    \left. -\frac{1}{8}{\cal P}_{n,n+2}{\cal P}_{n+1,n+3}
      -\frac{1}{8}{\cal P}_{n+1,n+2}{\cal P}_{n,n+3} \right),
\end{eqnarray}
with eigenvalues $-N/2$. The permutation operator expressed in terms of spin-1
operators reads 
\begin{equation}\label{eq:Pnm}
{\cal P}_{n,m} = {\bf S}_n \cdot {\bf S}_m 
    + \left({\bf S}_n \cdot {\bf S}_m\right)^{2} -1.
\end{equation}
Finite-$N$ data indicate that the trimer eigenstates are, in fact, ground
states.

\end{appendix}

%
%



\begin{thebibliography}{100}
\bibitem{note5}
The subtraction of constants in operators (\ref{Dl}) and (\ref{Tl}) removes
  artificial contributions $\propto \delta_{q,0}$ from the dimer and trimer
  structure factors, and thus makes them continuous at $q=0$.
\bibitem{note3}
Whether the singlet 2-magnon continuum observed in Fig.~2(c) really collapses
  into a number of $\propto N$ of degenerate excitations at $q=\pi$ cannot be
  decided on the basis of the recursion method. If the excitation at
  $\omega=\sqrt{10}J$ were degenerate, the recursion algorithm would pick out a
  particular direction in the subspace of these excitations such that $|\langle
  \lambda |A | G \rangle|^2 = \langle G|A^\dagger A | G \rangle$ and $\langle
  \lambda'| A |G\rangle=0$ for other (orthogonal) eigenvectors.
\bibitem{note7}
Bound states with positive interaction energy are a common occurrence in
  quantum systems on the lattice where $H$ and $-H$ are equally well defined.
\bibitem{note2}
Note of caution: States near the top or bottom of a continuum, which are more
  easily tracked than others, have transition rates which are likely to be
  affected by the singularity of the spectral-weight distribution at the band
  edge in the limit $N\to\infty$.
\bibitem{note10}
The $\theta$-dependence of $S_{SS}(\pi)$ has a maximum at $\theta > -\pi/4$ for
  finite $N$; as $N$ increases, the maximum grows and moves toward
  $\theta=-\pi/4$.
\bibitem{note11}
The $\theta$-dependence $S_{ZZ}(2\pi/3)$ has a maximum
  at $\theta < \pi/4$ for finite $N$; as $N$ increases, the maximum grows and
  moves toward $\theta=\pi/4$.
\bibitem{note4}
The vectors $|\Phi_k^D\rangle$, $k=1,2$ are non-orthogonal for finite $N$, but
  they become orthogonal as $N\rightarrow\infty$. The same is true for the
  vectors $|\Phi_k^T\rangle$, $k=1,2,3$.
\end{thebibliography}
\end{document}